\documentclass[a4paper, 12pt]{panl}

\usepackage[main=english,russian]{babel}
\usepackage{cite}
\usepackage{wrapfig}
\usepackage{graphicx}
\usepackage{amssymb}
\usepackage{amsfonts}
\usepackage{amsmath}
\usepackage{longtable}
\usepackage{rotating}
\usepackage{lscape}
\usepackage{epsfig}
\usepackage{multirow}
\usepackage{array}

\originalTeX

\begin{document}

\title{Search for dark matter mediator in the production of three and four top quarks.\\
}
\maketitle
\authors{E.\,Abasov $^{a, }$\footnote{E-mail: emil@abasov.ru},
E.\,Boos $^{a}$, V.\,Bunichev $^{a}$, P.\,Volkov $^{a}$, G.\,Vorotnikov $^{a}$,} 
\authors{L.\,Dudko $^{a}$, A.\,Zaborenko $^{a}$, E.\,Iudin $^{a}$, A.\,Markina $^{a}$, M.\,Perfilov $^{a}$, N.\,Savkova $^{b}$}
\setcounter{footnote}{0}

\from{$^{a}$\,Skobeltsyn Institute of Nuclear Physics, Lomonosov Moscow State University, Leninskie gory, GSP-1\\
Moscow 119991, Russian Federation\\}
\from{$^{b}$\,Sarov Branch of Lomonosov Moscow State University,\\
Sarov 607328, Russian Federation\\}
\begin{abstract}

\vspace{0.2cm}
In the context of simplified models of dark matter, the contributions of a scalar mediator to top quark production processes are considered. Tree-level and one-loop contributions of the mediator's decay into a top-antitop quark pair in two-, three-, and four-top quarks production processes are calculated. A significant contribution from diagrams involving the dark matter mediator is demonstrated in the total cross section for three- and four-top quark production processes, taking into account current experimental limits on model parameters. The perspective of searching for dark matter mediators in the processes under consideration is determined by the ability to reconstruct the final state with modern collider detectors and the experimental sensitivity that has already been achieved for such rare events.

\end{abstract}
\vspace*{6pt}


\label{sec:intro}
\section*{Introduction}
Convincing astrophysical and cosmological observations, such as studies of galaxy rotation curves~\cite{Rubin:1970zza} and gravitational lensing in galaxy cluster collisions~\cite{Harvey:2015hha}, indicate the presence of a large amount of hidden matter in the Universe, known as "dark matter" (DM). This dark matter can reveals itself through gravitational interaction and weak interactions with Standard Model (SM) fields. Based on the assumption that DM has a particle nature, various extensions of the SM are created to explain the origin of dark matter and its possible detection in experiments.

Direct searches for dark matter can be carried out through the interaction of cosmic DM particles with ground-based detectors~\cite{LUX:2012kmp}. At the same time, the Large Hadron Collider (LHC) provides a unique opportunity to observe and study the formation of DM particles under laboratory conditions. This opportunity is based on calculations of the DM particle annihilation cross-section, which is close to the weak interaction cross-section at LHC energies~\cite{PhysRevD.84.043510}. Since DM particles cannot be directly observed by LHC detectors, the main method of their detection is the joint production of dark sector particles and SM particles. In such a case, the DM particle in the final state carries away energy, which is observed as a "missing transverse energy" in the detector.

Possible interactions between DM particles and SM particles can occur through an intermediate particle known as a mediator. Depending on the interaction mechanism between the DM particles and SM particles, the mediator is endowed with various properties. One type of interaction could be a contact interaction involving a quark-antiquark pair or two gluons and two DM particles. In this case, the magnitude and distribution of the missing energy signal are determined by the nature and mass of the DM particles and the Lorentz structure of the interaction; the cross-section of the process is the main free parameter that can be measured. To model such scenarios, a basis of contact interaction operators can be introduced in effective field theory~\cite{GOODMAN2011185} to calculate the magnitude of possible signals.

Effective field theory is used under the assumption that the DM mediator is very heavy. In the case of a presumably lighter mediator, models that explicitly include mediators are necessary. Such models, known as simplified models~\cite{PhysRevD.79.075020, LHCNewPhysicsWorkingGroup:2011mji}, include particles and interactions beyond the SM. These simplified models~\cite{Abercrombie:2015wmb, Abdallah:2015ter} can be used in experimental searches for DM particles at the LHC. Although there are many different ways to construct simplified models, currently, those in which DM particles interact with SM particles via new scalar, pseudoscalar, or vector mediators appear theoretically attractive.

Under the assumption of minimal flavor violation (MFV)~\cite{Chivukula1987CompositetechnicolorSM, PhysRevLett.65.2939, BURAS2001161, DAMBROSIO2002155}, third-generation quarks can play a significant role~\cite{PhysRevD.88.063510} in such interactions. Therefore, it is particularly interesting to consider the processes of DM production at colliders in association with a top quark, as it has the largest mass. This idea has motivated experimental searches for events in which DM particles are produced in association with a single top quark ($t/\Bar{t}$ + DM) and a pair of top quarks ($t\Bar{t}$ + DM)~\cite{CMS:2019zzl, ATLAS:2022ygn}. No significant excess over SM predictions has been found in LHC experiments, and the issue of improving the analysis efficiency remains relevant.

In simplified DM models, it is assumed that dark matter particles $\chi$ are Dirac fermions interacting with the SM and DM sectors via either a massive electrically neutral scalar particle $\Phi$ (scalar mediator) or a pseudoscalar particle $A$ (pseudoscalar mediator)~\cite{Buckley:2014fba}. The interaction Lagrangians of scalar and pseudoscalar particles with SM and DM fermions are as follows:
$$
\begin{gathered}
L_{\Phi} = g_\chi \Phi \bar{\chi} \chi + \frac{g_f \Phi}{\sqrt{2}} \sum_f \left(y_f \bar{f} f \right) \\
L_A = i g_\chi A \bar{\chi} \gamma^5 \chi + i \frac{g_f A}{\sqrt{2}} \sum_f \left(y_f \bar{f} \gamma^5 f \right)
\end{gathered}
$$
Here, summation is done over all SM fermions, denoted as $f$. The parameters $y_f = \sqrt{2} m_f / v$ are Yukawa coupling constants with the vacuum expectation value of the Higgs field being 246 GeV. $g_\chi$ is the coupling constant of DM fermions with the mediator and $g_f$ is the coupling constant of SM fermions with the mediator. To reduce the number of model parameters, it is usually assumed that the value of $g_\chi$ is the same for all fermion flavors. Under the assumption of minimal flavor violation, such a simplified model contains a minimal set of four free parameters: the mass of dark matter particles $m_\chi$, the mass of the mediator $m_{\Phi/A}$, $g_f$, and $g_\chi$.

According to the recommendations of the LHC Dark Matter Working Group~\cite{Boveia:2016mrp}, in our study the coupling parameter values are taken to be $g_f = g_\chi = 1$, and the DM particle mass $m_\chi = 1$ GeV. For these parameter values, scalar and pseudoscalar mediator masses up to about 400 GeV are currently excluded~\cite{CMS:2021eha,ATLAS:2022ygn}. In most cases, the DM mediator decays into a pair of DM particles. With the above assumptions the branching ratio for the decay of a 400 GeV scalar mediator into a pair of DM particles is 84\%. However, the DM mediator can also decay into a pair of top and antitop quarks, with a branching ratio of 16\% for a 400 GeV scalar mediator. Accordingly, one can assume that the DM contribution to the production processes of top quark pairs, 3-top, and 4-top quarks can be quite significant. In this work, we explore the possibility of detecting DM in such processes within a simplified model with a scalar mediator. The pseudoscalar or vector mediator leads to almost the same or even better sensitivity to DM contribution.

The article is organized as follows. Section~\ref{sec:ttbar} analyzes the contribution of the dark matter mediator to the dominant top quark pair production process $pp \to \bar{t}t$. Section~\ref{sec:4top} investigates the process $pp \to \bar{t}t\bar{t}t$, and Section~\ref{sec:3topW} considers the production of three top quarks. Given the objective difficulties of distinguishing the four-top quark production process from the three-top quark process with next-to-leading-order (NLO) corrections, Section~\ref{sec:3top} examines the process $pp \to t\Bar{t}tW^-\Bar{b}$, which includes both three-top and four-top quark processes.

The modeling and numerical calculations were performed using the CompHEP4.6rc1~\cite{CompHEP:2004qpa, Pukhov:1999gg} and MadGraph5~\cite{Alwall:2014hca} computational packages. The models used in MadGraph5 are DMScalar, which includes an effective vertex $gg\Phi$~\cite{Pinna:2017tay}, and DMSimp, which facilitates NLO calculations~\cite{Backovic:2015soa}. All calculations are presented for proton-proton collision energies of 14 TeV for the LHC, with parton distribution functions being $\mathrm{NNPDF23\_nlo\_as\_0118}$ from the LHAPDF6 package~\cite{Buckley:2014ana}. Since at this stage of research it is only important to compare generators and estimate the possible contribution of DM diagrams, only Monte Carlo uncertainties are given for the cross-sections.

\section{Pair production of $t\Bar{t}$ in a Simplified DM Model with a Scalar Mediator}
\label{sec:ttbar}
The dominant process involving the vertex $\Phi \to t\Bar{t}$ at the LHC is the process $pp \to \Bar{t}t$. Therefore, it makes sense to start analyzing the contribution of the DM mediator with this process. The renormalization and factorization scale in the MC calculations for pair production is fixed at $m_t = 172.5$ GeV.

The leading order (LO) cross-section for top quark pair production $pp \to \Bar{t}t$ within the SM, calculated using the CompHEP software package, is $617.9 \pm 0.2$ pb for a top quark mass $m_t = 172.5$ GeV. Pair production of top quarks occurs either in the collision of a light quark-antiquark pair with the exchange of a virtual gluon or in the collision of two gluons from two colliding hadrons.

At the LHC, the process of $t\Bar{t}$ pair production in the collision of two gluons dominates, with a cross-section of $535.6 \pm 0.2$ pb.

In the framework of a simplified DM model with a scalar mediator, diagrams with the production of a DM mediator and its subsequent decay into a $t\Bar{t}$ pair also contribute to the processes of top quark pair production. The cross-section for the process $pp \to \Bar{t}t$ within this model, calculated using the CompHEP software package, is $618.26 \pm 0.06$ pb for a top quark mass $m_t = 172.5$ GeV and a scalar DM mediator mass $m_\Phi = 400$ GeV.

\begin{figure}[h]
    \centering
    \includegraphics[width=1.\linewidth]{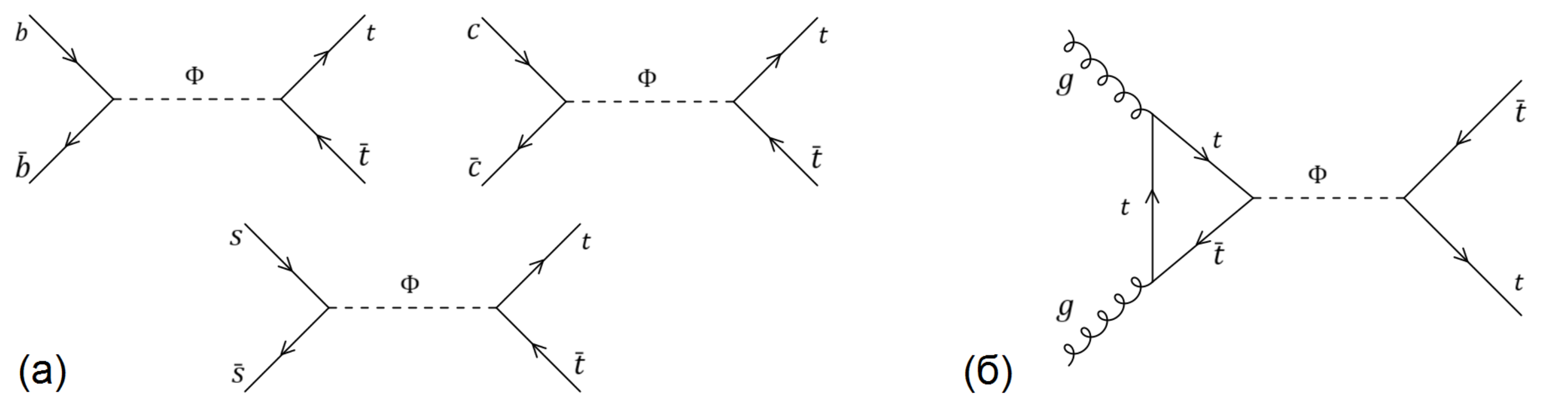}
    \caption{Feynman diagrams describing the production of a top-antitop quark pair in the collision of a light quark-antiquark pair (a) and in the collision of two gluons from two colliding hadrons (b) within the framework of a simplified DM model with a scalar mediator.}
    \label{fig:tt_diagrams_DM}
\end{figure}

The diagrams for the DM contribution to the subprocess $q\Bar{q} \to t\Bar{t}$ are shown in Fig.~\ref{fig:tt_diagrams_DM}(a). The cross-section for these processes is $(4.282 \pm 0.002) \cdot 10^{-3}$ pb, which is about 5 orders of magnitude smaller than the cross-section for the process $pp \to \Bar{t}t$ within the SM.

In the fusion of two gluons $gg \to t\Bar{t}$, the main diagram is the production of a scalar mediator and its subsequent decay into a top quark pair – this is a loop diagram, shown in Fig.~\ref{fig:tt_diagrams_DM}(b). The cross-section for this process, calculated with the same parameters in the CompHEP and MadGraph packages, is almost three orders of magnitude larger than the cross-section for processes with initial state quarks, being $0.917 \pm 0.025$ pb and $0.884 \pm 0.001$ pb, respectively. However, this cross-section is small compared to the total cross-section for the process $pp \to \Bar{t}t$ within the SM, which does not allow using this process for DM searches.
All the aforementioned cross-sections are presented in Table~\ref{tab:2top_cx}.

\begin{table}[ht]
\centering
\begin{tabular}{|c|c|}
    \hline
    Process & Cross-section (pb)\\
    \hline
    $pp \to t\Bar{t}$ (SM)$^{*}$ & $617.93 \pm 0.23$\\  
    \hline
    $gg \to t\Bar{t}$ (SM)$^{*}$ & $535.63 \pm 0.23$\\  
    \hline
    $pp \to t\Bar{t}$ (SM + DM)$^{*}$ & $618.26 \pm 0.06$\\  
    \hline
    $q\Bar{q} \to t\Bar{t}$ (tree diagrams DM)$^{*}$ & $(4.282 \pm 0.002) \cdot 10^{-3}$\\  
    \hline
    $gg \to t\Bar{t}$ (loop diagram DM)$^{*}$ & $0.917 \pm 0.025$\\ 
    \hline 
    $gg \to t\Bar{t}$ (loop diagram DM)$^{**}$ & $0.884 \pm 0.001$\\  
    \hline 
\end{tabular}
\caption{Cross-sections for top quark pair production processes within the SM and within a simplified DM model with a scalar mediator, calculated using CompHEP (${*}$) and MadGraph(${**}$).}
\label{tab:2top_cx}
\end{table}

Thus, the search for the production of a DM mediator decaying into a top-antitop quark pair in the processes of top quark pair production is unpromising.

\section{Four top quarks production $t\Bar{t}t\Bar{t}$ in a Simplified DM Model with a Scalar Mediator}
\label{sec:4top}

The production of four top quarks $pp \to t\Bar{t}t\Bar{t}$ is one of the rarest SM processes observable with current LHC parameters. The cross-section of this process within the SM, calculated using the MadGraph package, is $8.88 \pm 0.02$ fb. In this section, as well as in sections~\ref{sec:3topW} and~\ref{sec:3top}, the renormalization and factorization scales in the calculations are set to 350 GeV. The process $pp \to t\Bar{t}\Phi \to t\Bar{t}t\Bar{t}$ is one of the most promising channels for the production of a DM mediator~\cite{Craig:2016ygr, ATLAS:2022rws}.

\begin{figure}[!h!]
\centering
\includegraphics[width = .7\linewidth]{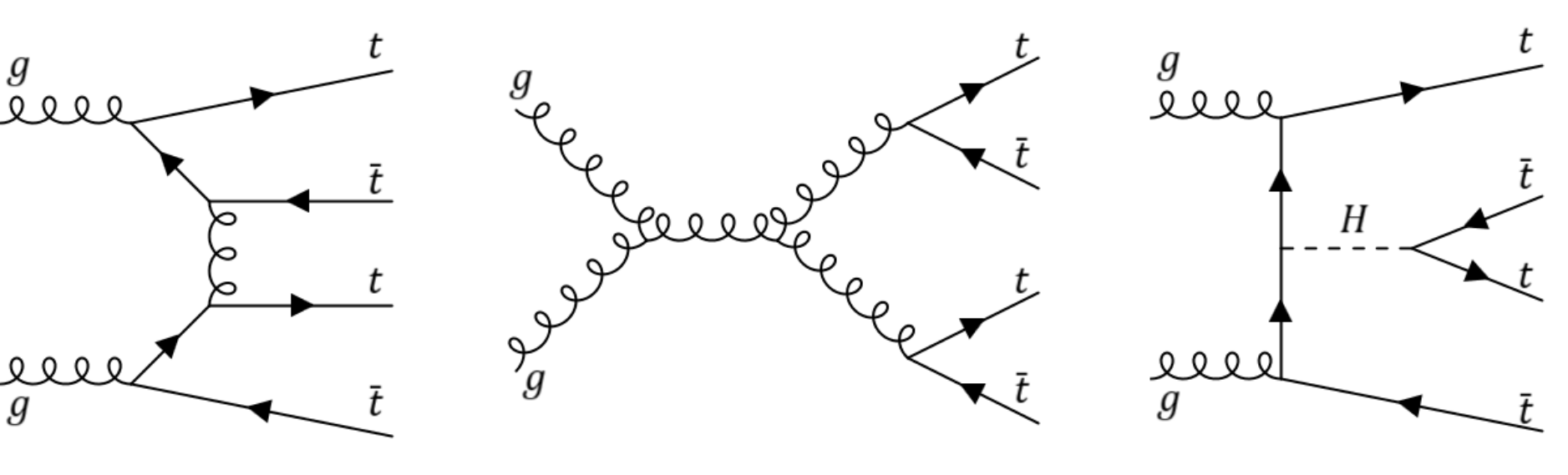}
\includegraphics[width = .6\linewidth]{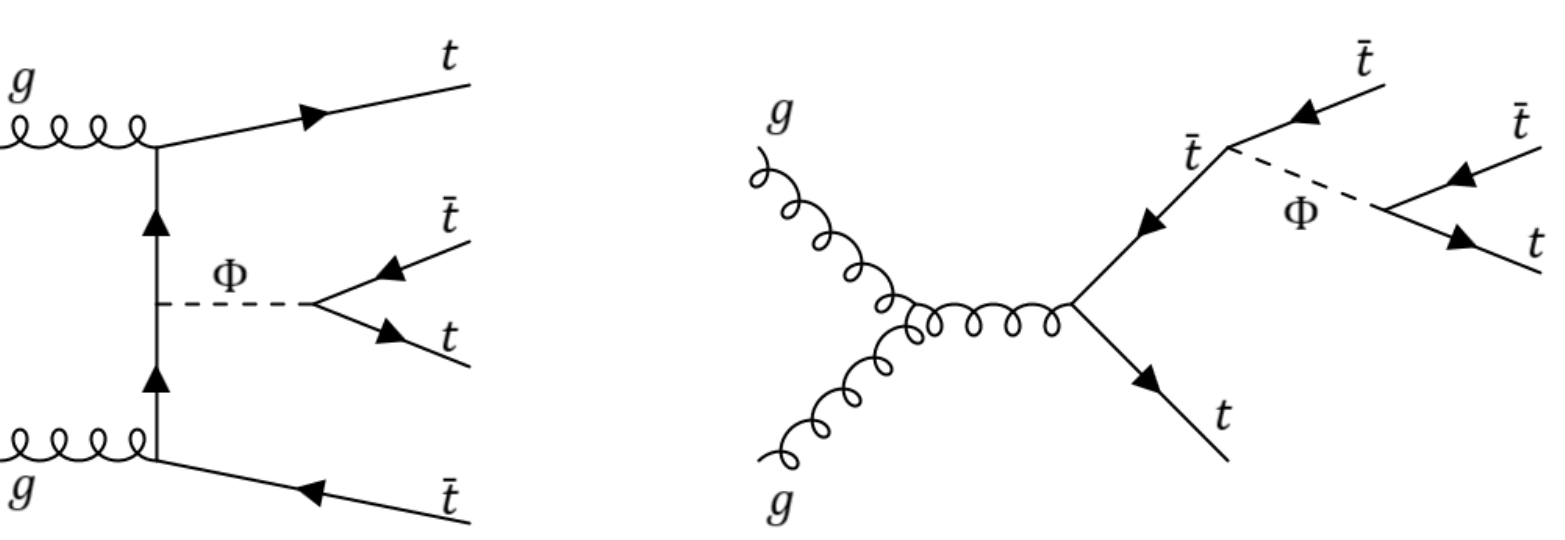}
\caption{Feynman diagrams describing the production of four top quarks in the collision of two gluons from two colliding hadrons. The top row shows representative diagrams within the SM, and the bottom row shows diagrams within the Simplified Dark Matter Model with a scalar mediator.}
\label{fig:4t_diagrams}
\end{figure}

The main contribution to the four top-quark production cross-section at the LHC comes from diagrams with gluons in the initial state. Representative diagrams are shown in Fig.\ref{fig:4t_diagrams}. In the framework of a simplified DM model with a scalar mediator, additional contributions come from diagrams involving the production of the DM mediator. Examples of such diagrams are also shown in Fig.\ref{fig:4t_diagrams}. Table~\ref{tab:4top_cx} presents the cross-sections of the process $gg \to t\Bar{t}t\Bar{t}$ within the SM and within the simplified DM model with a scalar mediator at leading order (LO) in perturbation theory, calculated in CompHEP and MadGraph. The cross-section of this process, calculated within the DM model, significantly exceeds the cross-section calculated within the SM.

Table~\ref{tab:4top_MG} presents the cross-section of the process $gg \to t\Bar{t}t\Bar{t}$ in the considered DM model, including loop diagrams, as well as the separate contributions of LO and loop diagrams involving the DM mediator, calculated in MadGraph. The contribution of diagrams involving the DM mediator to the cross-section of the process $gg \to t\Bar{t}t\Bar{t}$ within the simplified DM model is approximately 32\%, and searching for DM manifestations in four top-quark production processes may have good prospects. The contribution of loop diagrams is two orders of magnitude smaller than the contribution of other DM diagrams, and its influence on the total cross-section is negligible.
\begin{table}[h]
    \centering
    \resizebox{\textwidth}{!}{\begin{tabular}{|c|c|c|c|}
        \hline
        Process & CompHEP Cross-section (fb) & MadGraph Cross-section (fb)\\
        \hline
        $gg \to t\Bar{t}t\Bar{t}$ (SM) & $7.71 \pm 7.55*10^{-3}$  & $7.79 \pm 2.3*10^{-2}$\\  
        \hline
        $gg \to t\Bar{t}t\Bar{t}$ (SM + DM) & $11.15 \pm 1.60*10^{-2}$  & $11.41 \pm 3.4*10^{-2}$\\  
        \hline
        
    \end{tabular}}
    \caption{LO Cross-sections of the process $gg \to t\Bar{t}t\Bar{t}$ within the SM and within the simplified DM model with a scalar mediator, calculated in CompHEP and MadGraph}
    \label{tab:4top_cx}
\end{table}
\begin{table}[!h!]
    \centering
    \resizebox{\textwidth}{!}{\begin{tabular}{|m{18em}|c|c|}
        \hline
        Description of the process $gg \to t\Bar{t}t\Bar{t}$ (DM) & MadGraph Cross-section (fb)\\
        \hline
        Full set of diagrams, including loop diagrams & $11.44 \pm 3*10^{-2}$\\
        \hline
        Contribution of diagrams with a mediator, excluding loop diagrams & $2.89 \pm 9.9*10^{-3}$\\
        \hline
        Contribution of loop diagrams with a mediator & $8.67*10^{-2} \pm 2.74*10^{-4}$\\
        \hline
    \end{tabular}}
    \caption{Cross-section of the process $gg \to t\Bar{t}t\Bar{t}$ within the simplified DM model with a scalar mediator, including loop diagrams; contribution of diagrams with a mediator, with and without considering loop diagrams.}
    \label{tab:4top_MG}
\end{table}
\section{Processes with the final state $t\Bar{t}tW^-$ within the framework of a simplified DM model with a scalar mediator}
\label{sec:3topW}
The process of producing three top quarks in association with a W boson is an even rarer process than the production of four top quarks and has not yet been experimentally observed.

For the dominant processes of three top-quark production within the SM, it has been shown that the interference of QCD and electroweak diagrams is quite large and negative, so it is important to consider the full set of diagrams to calculate the cross-sections of these processes~\cite{Boos:2021yat}. A comparison of the contributions is given in Table~\ref{tab:3top_SM}.

Table~\ref{tab:3topW_MG} describes the results of calculations of various DM contributions to this process. It is noticeable that the DM contribution is significantly higher than for the production of four top quarks, with the cross-section of diagrams involving the DM mediator accounting for 60\% of the total cross-section. As in the production of four top quarks, loop diagrams give cross-sections several orders of magnitude smaller and can be neglected in this process. Thus, this process is highly promising for DM searches at colliders.
\begin{table}[!h!]
    \centering
    \resizebox{\textwidth}{!}{\begin{tabular}{|m{18em}|c|c|}
        \hline
        Process $pp \to t\Bar{t}tW^-$ & MadGraph Cross-section (fb)\\
        \hline
        Full set of diagrams in the SM & $0.62 \pm 1.6*10^{-3}$\\  
        \hline
        Full set of diagrams in the DM model, excluding loop diagrams & $1.54 \pm 5*10^{-3}$\\  
        \hline
        Full set of diagrams in the DM model, including loop diagrams & $1.53 \pm 5*10^{-3}$\\  
        \hline
        Contribution of diagrams with a DM mediator, excluding loop diagrams & $0.90 \pm 2.4*10^{-3}$\\  
        \hline
        Contribution of loop diagrams with a DM mediator & $1.20*10^{-3} \pm 3.7*10^{-6}$\\  
        \hline
    \end{tabular}}
    \caption{Cross-sections of the process $pp \to t\Bar{t}tW^-$ and individual contributions within the SM and the Simplified DM Model with a Scalar Mediator.}
    \label{tab:3topW_MG}
\end{table}

\section{Processes with the final state $t\Bar{t}tW^-\Bar{b}$ within the framework of a simplified DM model with a scalar mediator}
\label{sec:3top}
The processes of producing three top quarks in association with a W boson at NLO and the processes of producing four top quarks with the subsequent decay of one top quark share the same final state $t\Bar{t}tW^-\Bar{b}$. Separating these contributions is analogous to the task of distinguishing contributions from $tWb$ and $t\Bar{t}$ processes~\cite{Boos:2023kpp}, requiring detailed study due to significant negative interference between these processes. Studying the complete set of diagrams for the process $pp \to t\Bar{t}tW^-\Bar{b}$ can provide insight into the DM contributions to both processes without the need for their separate consideration.

Table~\ref{tab:3top_MG} presents the cross-sections of the process $gg \to t\Bar{t}tW^-\Bar{b}$ calculated within the SM and the considered DM model, with and without loop diagrams included. Specifically calculated and presented are the contributions of diagrams involving the DM mediator without loop diagrams and the contribution of loop diagrams. The contribution of diagrams involving the DM mediator to the cross-section of $gg \to t\Bar{t}tW^-\Bar{b}$ is comparable to that of $gg \to t\Bar{t}t\Bar{t}$ due to its dominant contribution (Table~\ref{tab:3top_SM}), approximately 33\%.

\begin{table}[!h!]
    \centering
    \resizebox{\textwidth}{!}{\begin{tabular}{|m{18em}|c|c|}
        \hline
        Process $gg \to t\Bar{t}tW^-\Bar{b}$ & MadGraph Cross-section (fb)\\
        \hline
        Full set of diagrams in the SM & $15.55 \pm 4.7*10^{-2}$\\  
        \hline
        Full set of diagrams in the DM model, excluding loop diagrams & $23.02 \pm 6.2*10^{-2}$\\  
        \hline
        Full set of diagrams in the DM model, including loop diagrams & $23.16 \pm 6*10^{-2}$\\  
        \hline
        Contribution of diagrams with a DM mediator, excluding loop diagrams & $6.03 \pm 1.2*10^{-2}$\\  
        \hline
        Contribution of loop diagrams with a DM mediator & $0.17 \pm 5*10^{-4}$\\  
        \hline
    \end{tabular}}
    \caption{Cross-sections of the process $gg \to t\Bar{t}tW^-\Bar{b}$ and individual contributions within the SM and the Simplified DM Model with a Scalar Mediator.}
    \label{tab:3top_MG}
\end{table}

\begin{table}[!h!]
    \centering
    \begin{tabular}{|c|c|c|}
        \hline
        Process Description & MadGraph Cross-section (fb)\\
        \hline
        $pp \to t\Bar{t}tW^-$, QCD+EW & $0.622 \pm 1.6*10^{-3}$\\  
        \hline
        $pp \to t\Bar{t}tW^-$, QCD & $0.453 \pm 1.3*10^{-3}$\\  
        \hline
        $pp \to t\Bar{t}tW^-$, EW & $0.452 \pm 1.3*10^{-3}$\\  
        \hline
        $pp \to t\Bar{t}tW^-$, QCD+EW+DM & $1.54 \pm 5*10^{-3}$\\  
        \hline
        \hline
        $gg \to t\Bar{t}tW^-\Bar{b}$, QCD+EW & $15.5 \pm 4.7*10^{-2}$\\  
        \hline
        $gg \to t\Bar{t}tW^-\Bar{b}$, QCD & $14.0 \pm 4.6*10^{-2}$\\  
        \hline
        $gg \to t\Bar{t}tW^-\Bar{b}$, EW & $6.08 \pm 1.4*10^{-2}$\\  
        \hline
        $gg \to t\Bar{t}tW^-\Bar{b}$, QCD+EW+DM & $23.02 \pm 6*10^{-2}$\\  
        \hline
    \end{tabular}
    \caption{Comparison of QCD and electroweak contributions to processes involving the production of 3 and 4 top quarks. Notations are introduced: QCD - contribution from QCD diagrams only, EW - contribution from electroweak diagrams (QED==3), QCD+EW - full set of diagrams within the SM (QED=3), QCD+EW+DM - full set of diagrams in the DM model, excluding loop diagrams.}
    \label{tab:3top_SM}
\end{table}

\section{Results}
\label{sec:conclusion}
The possibility of searching for a dark matter (DM) mediator decaying into a pair of top-antitop quarks in processes involving the production of two, three, and four top quarks has been considered. Taking into account modern experimental constraints and the parameter space recommended for calculations in the simplified model with a scalar mediator, potential DM contributions in tree-level and loop-level approximations were computed for the primary processes accessible to contemporary collider experiments. It has been demonstrated that for the dominant process of top-antitop quark pair production, the possible contribution from the DM mediator, including loop diagrams, amounts to approximately 0.1\% and is not of interest for experimental searches.

Conversely, significantly rarer processes involving the production of four and three top quarks exhibit substantial potential, where the DM contribution can reach 32\% and 60\% of the total cross-section, respectively, for the recommended parameters of the simplified DM model. Considering the complexity of experimentally distinguishing between the processes involving three and four top quarks, the process $pp \to t\Bar{t}tW^-\Bar{b}$ was computed, encompassing the production of both three and four top quarks and the resulting interferences. It has been shown that the DM contribution can constitute 33\% of the total cross-section of this process, making it the most promising from the standpoint of experimental searches in modern collider experiments.

Similar sensitivity can be expected for simplified models with pseudoscalar and vector mediators.

\section*{Acknowledgments}
\label{sec:acknowledgement}
This study was conducted within the scientific program of the Russian National Center for Physics and Mathematics, section 5 «Particle Physics and Cosmology».

\bibliographystyle{pepan}
\bibliography{main}

\end{document}